# An Analysis of the Heston Stochastic Volatility Model:

# Implementation and Calibration using Matlab[*]

Ricardo Crisóstomo[†]

December 2014

## Abstract

This paper analyses the implementation and calibration of the Heston Stochastic Volatility Model. We first explain how characteristic functions can be used to estimate option prices. Then we consider the implementation of the Heston model, showing that relatively simple solutions can lead to fast and accurate vanilla option prices. We also perform several calibration tests, using both local and global optimization. Our analyses show that straightforward setups deliver good calibration results. All calculations are carried out in Matlab and numerical examples are included in the paper to facilitate the understanding of mathematical concepts.

[*] The author acknowledges the comments of seminar participants at the CNMV. A previous version of this work was circulated in partial fulfillment of the requirements of an MSc degree in mathematics at the National Distance Education University (UNED). The opinions in this article are the sole responsibility of the author and they do not necessarily coincide with those of the CNMV.
[†] Comisión Nacional del Mercado de Valores (CNMV). c/ Edison 4, 28006 Madrid. Email: rcayala@cnmv.es.

# List of Contents

# 1. Introduction

The Black and Scholes (BSM) model provides a coherent framework for pricing European options. However, this method is based on several assumptions that are not representative of the real world. In particular, the BSM model assumes that volatility is deterministic and remains constant through the option's life, which clearly contradicts the behavior observed in financial markets. While the BSM framework can be adapted to obtain reasonable prices for plain vanilla options, the constant volatility assumption may lead to significant mispricings when used to evaluate options with non-conventional or exotics features.

During the last decades several alternatives have been proposed to improve volatility modelling in the context of derivatives pricing. One of such approaches is to model volatility as a stochastic quantity. By introducing uncertainty in the behavior of volatility, the evolution of financial assets can be estimated more realistically. In addition, using appropriate parameters, stochastic volatility models can be calibrated to reproduce the market prices of liquid options and other derivatives contracts.

One of the most widely used stochastic volatility model was proposed by Heston in 1993. The Heston model introduces a dynamic for the underlying asset which can take into account the asymmetry and excess kurtosis that are typically observed in financial assets returns. It also provides a closed-form valuation formula that can be used to efficiently price plain vanilla options. This will be particularly useful in the calibration process, where many option repricings are usually required in order to find the optimal parameters that reproduce market prices.

In this paper we analyze the valuation of financial options using the Heston model. Our aim is to illustrate the use of the model with an emphasis on the implementation and calibration. Section 2 presents the valuation framework and explains how characteristic functions can be used to estimate option prices. Section 3 introduces the Heston model and discusses the implementation of its closed-form solution. Finally, Section 4 analyzes the calibration problem, considering both local and global optimization methods.

For all relevant sections, generic and ready-to-use Matlab codes have been developed and numerical examples are provided in order to illustrate the use of the Matlab routines.

## 2. From Characteristic Functions to Option Prices

A considerable amount of research has been recently devoted to analyze the use of characteristic functions in option's valuation. The rationale is that when you go beyond the classical BSM framework, the underlying stochastic processes that are used to calculate option values have characteristic functions which are simpler and more tractable than their density functions. Therefore, in many sophisticated models, it is easier to work with characteristics functions instead of using density functions.

### 2.1 The General Valuation Framework

When markets are complete and arbitrage-free, option values can be calculated as the present value of their expected payoff under the risk-neutral measure

$$V_0 = e^{-rT} E_Q \left[ H(S_t) \right] \qquad (2.1)$$

where $V_0$ is the option value at time $t$ = 0, $r$ is the risk free rate, $T$ is the time to maturity and $H(S_t)$ is the option payoff. In order to use (2.1), we first need to specify the dynamics of the price process $S_t$. In particular, since we are working with expectations, we should consider the probability distribution of $S_t$ at (potentially) different times, as required by each option payoff.

In the classical framework, the expectation above is obtained by means of the risk-neutral density. For instance, the payoff of a European call with strike $K$ and expiration date $T$ is given by $H(S_t) = (S_T - K)^+$. Consequently, its value at time $t$ = 0 is

$$C_0 = e^{-rT} \int_0^\infty (S_T - K)^+ q(S_T) dS_T \qquad (2.2)$$

where $q(S_T)$ is the risk-neutral density of the underlying asset $S_t$ at the terminal date $T$. The problem with (2.2) is that there are many price processes for which the density function $q(S_T)$ is not available in a closed-form or is difficult to obtain. However, if we work with the logarithm of the underlying asset price, there are many of such price processes with both simpler and analytically tractable characteristic functions.

Characteristic functions exhibit a one-to-one relationship with density functions. In particular, the characteristic function of a given stochastic process $X$, is the Fourier transform of its probability density function

$$\psi(w) = E[e^{iwX}] = \int_{-\infty}^\infty e^{iwx} f(x) dx \qquad (2.3)$$

Therefore, by applying the Fourier Inversion theorem, we can recover the density function of the process $X$ in terms of its characteristic function

$$f(x) = \frac{1}{2\pi} \int_{-\infty}^\infty e^{-iwx} \psi(w) dw \qquad (2.4)$$

Given this relationship, all the probability evaluations that are required to calculate options values can be also computed using characteristic functions.

## 2.2 Valuing a European Call through Characteristic Functions

Following the reasoning in Heston (1993), the value of a European call option can be obtained by using a probabilistic approach

$$C_0 = S_0 \, \Pi_1 - e^{-rT} K \, \Pi_2 \qquad (2.5)$$

where $\Pi_1$ and $\Pi_2$ are two probability-related quantities. Specifically, $\Pi_1$ is the option delta and $\Pi_2$ is the risk-neutral probability of exercise $P(S_T > K)$. Instead of using density functions, these probabilities can be computed via characteristic functions as follows (proof in Appendix A):

$$\Pi_1 = \frac{1}{2} + \frac{1}{\pi} \int_0^\infty \text{Re} \left[ \frac{e^{-iw\ln(K)} \psi_{\ln S_T}(w-i)}{iw \psi_{\ln S_T}(-i)} \right] dw \qquad (2.6)$$

$$\Pi_2 = \frac{1}{2} + \frac{1}{\pi} \int_0^\infty \text{Re} \left[ \frac{e^{-iw\ln(K)} \psi_{\ln S_T}(w)}{iw} \right] dw \qquad (2.7)$$

Therefore, starting with the characteristic function of the log-price $\psi_{\ln S_T}(w)$, we can estimate the price of a European call option by first calculating the probabilities (2.6) and (2.7) and then substituting their values in (2.5). This method presents two main advantages:

- **Generality**: This approach can be applied for any underlying price process $S_t$ whose characteristic function is known.
- **Semi-analytical solution:** The integrands in (2.6) and (2.7) should be evaluated numerically. However, they are smooth functions that decay rapidly and can be evaluated efficiently using appropriate integration routines[1]. This lead to numerical implementations that can value plain vanilla options in a fraction of a second[2].

## 2.3 An Application to the Black and Scholes Model

Before moving into the Heston model, we will apply the characteristic function method to value a call option under the BSM framework. The risk-neutral dynamics of the underlying asset in BSM are described by a Geometric Brownian Motion

$$dS_t = rS_t dt + \sigma S_t dW_t \qquad (2.8)$$

---

[1] See Kahl and Jäckel (2005) or Schmelzle (2010).
[2] For example, using the Matlab's implementation proposed in this paper, the computational times required for pricing a European call option are 0.003087 seconds in the BSM model and 0.004866 seconds in the Heston model.

where $S_t$ is the price of the underlying asset at time $t$, $r$ is the risk free rate, $\sigma$ is the volatility of the underlying returns, and $W_t$ is a Weiner process. Using stochastic calculus, equation (2.8) can be easily solved to yield

$$S_t = S_0 e^{(r - \frac{1}{2}\sigma^2)t + \sigma\sqrt{t}\,Z} \qquad (2.9)$$

where $Z$ is the standard normal distribution. Therefore, the distribution of $S_t$ is lognormal, while $\ln(S_t)$ is normally distributed. In particular, the risk-neutral evolution of $\ln(S_t)$ is normally distributed with mean $\ln(S_0) + (r - 0.5\sigma^2)t$ and variance $\sigma^2 t$. This means that, in practice, it is easier to work with the process $\ln(S_t)$ rather than using $S_t$ directly.

**Black and Scholes Characteristic Function**

The characteristic function of a normal random variable is given by

$$\psi(w) = e^{iw(\text{mean}) - \frac{1}{2}w^2(\text{variance})} \qquad (2.10)$$

Therefore, the characteristic function of $\ln(S_t)$ can be easily calculated as

$$\psi_{\ln(S_t)}^{BSM}(w) = e^{iw[\ln(S_0) + (r - 0.5\sigma^2)t] - 0.5w^2\sigma^2 t} \qquad (2.11)$$

Once we have the characteristic function, the next step is to estimate $\Pi_1$ and $\Pi_2$. These probabilities can be computed by numerical integration or, alternatively, Euler's formula ($e^{ix} = \cos x + i\sin x$) could be applied to further expand (2.6) and (2.7), and in order to obtain more specific expressions for $\Pi_1$ and $\Pi_2$ under the BSM framework.

Since our aim is to gain a better understanding of the general characteristic function approach, we will compute $\Pi_1$ and $\Pi_2$ directly using (2.6) and (2.7). We will repeat this procedure in section 3, where we will use $\psi_{\ln(S_t)}^{Heston}(w)$ instead of $\psi_{\ln(S_t)}^{BSM}(w)$ in order to calculate the value of a European call under the Heston model.

Function 1 below (*chfun_norm.m*) shows how to compute the characteristic function of the BSM model in Matlab, while function 2 (*call_bsm_cf.m*) calculates the call value based on equations (2.5) to (2.7). In addition, example 1 illustrates the practical use of these functions by pricing an individual call option. As a reference, within the BSM framework, the estimated value of a call option with parameters $S_0 = 100$, $K = 100$, $\sigma = 0.15$, $r = 0.02$ and $t = T = 1$ is $C_0 = 8.9160$. As the example shows, using the characteristic function approach, we obtain the same call value.

**Matlab Function 1: Characteristic function of the Black-Scholes model (chfun_norm.m)**

```
function y = chfun_norm(s0, v, r, t, w)

%  Characteristic function of BSM.
%  y = chfun_norm(s0, v, r, t, w)

%  Inputs:
%  s0: stock price
%  v: volatility
```

```
%  r: risk-free rate
%  t: time to maturity
%  w: points at which to evaluate the function

mean =log(s0)+ (r-v^2/2)*t;          % mean
var = v^2*t;                          % variance
y = exp((i.*w*mean)-(w.*w*var*.5));   % characteristic function of log (St) evaluated at points w
end
```

**Matlab Function 2: Call value in the Black-Scholes model (call_bsm_cf.m)**

```
function y = call_bsm_cf(s0, v, r, t, k)

%  BSM call value calculated using formulas 2.5 to 2.7
%  y = call_bsm_cf(s0, k, v, r, t, w )

%  Inputs:
%  s0: stock price
%  v: volatility
%  r: risk-free rate
%  t: time to maturity
%  k: option strike
%  chfun_norm: Black-Scholes characteristic function

%  1st step: calculate pi1 and pi2
   % Inner integral 1
int1 = @(w,s0,v,r,t,k) real(exp(-i.*w*log(k)).*chfun_norm(s0,v,r,t,w-i)./(i*w.*chfun_norm(s0, v, r, t, -i)));
int1 = integral(@(w)int1(w,s0,v,r,t,k),0,100); %numerical integration
pi1 = int1/pi+0.5;

   % Inner integral 2
int2 = @(w,s0,v,r,t,k) real(exp(-i.*w*log(k)).*chfun_norm(s0, v, r, t, w)./(i*w));
int2 = integral(@(w)int2(w,s0, v, r, t, k),0,100); %numerical integration
pi2 = int2/pi+0.5; % final pi2

%  2nd step: calculate call value
y = s0*pi1-exp(-r*t)*k*pi2;
end
```

**Numerical Example 1: Call option valuation using the Black-Scholes model**

```
% function y = call_bsm_cf(s0, v, r, t, k)

>> call_bsm_cf(100, 0.20, 0.02, 1, 100)

ans =  8.9160
```

## 3. The Heston Model

In 1993, Heston proposed a stochastic volatility model where the underlying asset behavior was characterized by the following risk-neutral dynamics

$$dS_t = rS_t dt + \sqrt{V_t} S_t dW_t^1$$
$$dV_t = a(\overline{V} - V_t)dt + \eta \sqrt{V_t} dW_t^2 \qquad (3.1)$$
$$dW_t^1 dW_t^2 = \rho dt$$

The parameters used in the model are the following:

- $S_t$ is the price of the underlying asset at time t
- $r$ is the risk free rate
- $V_t$ is the variance at time t
- $\overline{V}$ is the long-term variance
- $a$ is the variance mean-reversion speed
- $\eta$ is the volatility of the variance process
- $dW_t^1$, $dW_t^2$ are two correlated Weiner processes, with correlation coefficient $\rho$

Therefore, under the Heston model, the underlying asset follows an evolution process which is similar to the BSM model, but it also introduces a stochastic behavior for the volatility process. In particular, Heston makes the assumption that the asset variance $V_t$ follows a mean reverting Cox-Ingersoll-Ross process.

Stochastic volatility models tackle one of the most restrictive hypotheses of the BSM model; namely, the assumption that volatility remains constant during the option´s life. Observing financial markets it can be easily seen that volatility is not a constant quantity. This is also reflected in the different implied volatility levels at which options with different strikes and maturities trade in the market, which collectively give rise to the so-called volatility surface.

Among volatility models, Heston's dynamics exhibit several desirable properties. First, it models volatility as a mean-reverting process. This assumption is consistent with the behavior observed in financial markets. If volatility were not mean-reverting, markets would be characterized by a considerable amount of assets with volatility exploding or going near zero. In practice, however, these cases are quite rare and generally short-lived.

Second, it also introduces correlated shocks between asset returns and volatility. This assumption allows modelling the statistical dependence between the underlying asset and its volatility, which is a prominent feature of financial markets. For instance, in equity markets, volatility tends to increase when there are high drops in equity prices, and this relationship may have a substantial impact in the price of contingent claims.

Consequently, the Heston model provides a versatile modelling framework that can accommodate many of the specific characteristics that are typically observed in the behavior of financial assets. In particular, the parameter $\eta$ controls the kurtosis of the underlying asset return distribution, while $\rho$ sets its asymmetry.

However, as expected, these benefits come at the expense of higher complexity. Compared with BSM, the implementation of the Heston model requires more sophisticated mathematics and it also involves a more challenging process to calibrate the model to fit market prices.

## 3.1 Closed-form Solution of the Heston Model

One of the main advantages of the Heston model is that the price of European options can be estimated using a quasi-closed form valuation formula.

The development of the Heston formula follows the general approach that we explained in section 2. As we mentioned, the present value of a European call option can be estimated using a probabilistic approach

$$C_0 = S_0 \, \Pi_1 - e^{-rT} K \, \Pi_2 \quad (3.2)$$

where $\Pi_1$ and $\Pi_2$ are two probability-related quantities. Therefore, the call value under the Heston model can computed by first obtaining $\Pi_1$ and $\Pi_2$ using the dynamics described in (3.1) and then substituting their values in equation (3.2). However, the difficulty arises when we try to calculate these probabilities under the Heston dynamics, since the transition densities for this model are not available in a closed-form. Alternatively, as we showed earlier, $\Pi_1$ and $\Pi_2$ can also be obtained using characteristic functions.

## Heston Characteristic Function

In this section we start with the Heston characteristic function proposed by Gatheral (2006), but we also introduce an additional modification. In particular, the characteristic function that we will use through the paper is the following:

$$\psi_{\ln(S_t)}^{Heston}(w) = e^{[C(t,w)\bar{V} + D(t,w)V_0 + iw\ln(S_0 e^{rT})]}$$

$$C(t,w) = a \left[ r_- \cdot t - \frac{2}{\eta^2} ln\left( \frac{1 - ge^{-ht}}{1 - g} \right) \right]$$

$$D(t,w) = r_- \frac{1 - e^{-ht}}{1 - ge^{-ht}}$$

$$r_\pm = \frac{\beta \pm h}{\eta^2}; \quad h = \sqrt{\beta^2 - 4\alpha\gamma} \quad\quad (3.3)$$

$$g = \frac{r_-}{r_+}$$

$$\alpha = -\frac{w^2}{2} - \frac{iw}{2}; \quad \beta = a - \rho\eta iw; \quad \gamma = \frac{\eta^2}{2}$$

Our approach differs from Gatheral (2006) in that we apply the characteristic function method based on the process $\ln(S_t)$, instead of $\ln(S_t / K)$. Using this approach we obtain an expression for $\psi_{\ln(S_t)}^{Heston}(w)$ that can be directly used within the general pricing framework presented in section 2. This is in contrast with the formulation used in Heston (1993) and later in Gatheral (2006), where two distinct functions are used to calculate $\Pi_1$ and $\Pi_2$. Appendix B shows the equivalence of our approach to the methodology provided by Gatheral (2006)

It should be noted that the characteristic function presented in (3.3) already incorporates the risk-neutral behavior of the process $\ln(S_t)$. A discussion of the risk neutral paradigm in the Heston model is included in Appendix C.

### 3.2 Model Implementation

Although $\psi_{\ln(S_t)}^{Heston}(w)$ may have a complicated appearance, its implementation is quite straightforward. In particular, once we have estimated appropriate values for the model parameters $\{V_0, \bar{V}, a, \eta, \rho\}$, the Heston characteristic function can be easily evaluated using numerical software. Function 3 (*chfun_heston.m*) shows how to compute the Heston characteristic function in Matlab.

After obtaining $\psi_{\ln(S_t)}^{Heston}(w)$, the characteristic function can be substituted in (2.6) and (2.7) to calculate $\Pi_1$ and $\Pi_2$. Using these probabilities, equation (3.2) will provide the estimated value of a European call under the Heston Model. Function 4 (*call_heston_cf.m*) performs the calculations based on such equations.

Example 2 illustrates how to use these functions to value a call option where $S_0 = 1$, $K = 2$, $V_0 = 0.16$, $\bar{V} = 0.16$, $a = 1$, $\eta = 2$, $\rho = -0.8$ and $t = T = 10$. Kahl and Jäckel (2005) showed that the estimated value for this option under the Heston model is $C_0 = 0.0495$. As the example shows, our implementation yields the same call value.

It is also relevant to note that some authors compute the price of vanilla options in the Heston model using the Fast Fourier Transformation (FFT). This approach has the advantage that it can provide simultaneously the prices of options with different strikes and, therefore, it employs lower computational time[3]. However, the FFT approach introduces an additional parameter and its implementation requires modifying the general valuation formulas presented in section 2. Consequently, since our aim is to develop practical intuition on the Heston model, we will not employ this approach.

**Matlab Function 3: Characteristic function of the Heston model (chfun_heston.m )**

```
function y = chfun_heston(s0, v0, vbar, a, vvol, r, rho, t, w);

%  Heston characteristic function.
%  Inputs:
%  s0: stock price
%  v0: initial volatility (v0^2 initial variance)
%  vbar: long-term variance mean
%  a: variance mean-reversion speed
%  vvol: volatility of the variance process
%  r : risk-free rate
%  rho: correlation between the Weiner processes for the stock price and its variance
%  w: points at which to evaluate the function
%  Output:
```

---

[3] See Carr and Madam (1998).

```
%  Characteristic function of log (St) in the Heston model

%  Interim calculations
alpha = -w.*w/2 - i*w/2;
beta = a - rho*vvol*i*w;
gamma = vvol*vvol/2;
h = sqrt(beta.*beta - 4*alpha*gamma);
rplus = (beta + h)/vvol/vvol;
rminus = (beta - h)/vvol/vvol;
g=rminus./rplus;

%  Required inputs for the characteristic function
C = a * (rminus * t - (2 / vvol^2) .* log((1 - g .* exp(-h*t))./(1-g)));
D = rminus .* (1 - exp(-h * t))./(1 - g .* exp(-h*t));

%  Characteristic function evaluated at points w
y = exp(C*vbar + D*v0 + i*w*log(s0*exp(r*t)));
```

**Matlab Function 4: Call price in the Heston model (call_heston_cf.m)**

```
function y = call_heston_cf(s0, v0, vbar, a, vvol, r, rho, t, k)

%  Heston call value using characteristic functions.
%  y = call_heston_cf(s0, v0, vbar, a, vvol, r, rho, t, k)

%  Inputs:
%  s0: stock price
%  v0: initial volatility (v0^2 initial variance)
%  vbar: long-term variance mean
%  a: variance mean-reversion speed
%  vvol: volatility of the variance process
%  r: risk-free rate
%  rho: correlation between the Weiner processes of the stock price and its variance
%  t: time to maturity
%  k: option strike
%  chfun_heston: Heston characteristic function

%  1st step: calculate pi1 and pi2
  %  Inner integral 1
int1 = @(w, s0, v0, vbar, a, vvol, r, rho, t, k) real(exp(-i.*w*log(k)).*chfun_heston(s0, v0, vbar, a, vvol, r,
rho, t, w-i)./(i*w.*chfun_heston(s0, v0, vbar, a, vvol, r, rho, t, -i))); % inner integral1
int1 = integral(@(w)int1(w,s0, v0, vbar, a, vvol, r, rho, t, k),0,100); % numerical integration
pi1 = int1/pi+0.5; % final pi1

  %  Inner integral 2:
int2 = @(w, s0, v0, vbar, a, vvol, r, rho, t, k) real(exp(-i.*w*log(k)).*chfun_heston(s0, v0, vbar, a, vvol, r,
rho, t, w)./(i*w));
int2 = integral(@(w)int2(w,s0, v0, vbar, a, vvol, r, rho, t, k),0,100);int2 = real(int2);
pi2 = int2/pi+0.5; % final pi2
```

```
%  2rd step: calculate call value
y = s0*pi1-exp(-r*t)*k*pi2;
end
```

**Numerical Example 2: Call valuation in the Heston model.**

```
% function y = call_heston_cf(s0, v0, vbar, a, vvol, r, rho, t);

>> call_heston_cf(1, 0.16, 0.16, 1, 2, 0, -0.8, 10, 2)

ans =  0.0495
```

## 4. Calibration to Market Prices

Before using a pricing model we should ensure that it can produce accurate results for the options that are already traded in the market. Availability of closed-form solutions is particularly useful in the calibration process. Typically, when we seek to obtain the optimal model parameters that are able to reproduce market prices, we need to perform a substantial number of plain vanilla options repricings. Consequently, accurate and efficient pricing formulas are required in order to obtain reliable results within a reasonable timeframe.

### 4.1 Calibration Procedure in the Heston Model

The goal of calibration is to find the parameter set that minimizes the distance between model predictions and observed market prices. In particular, using the risk-neutral measure, the Heston model has five unknown parameters $\Omega = \left\{ V_0, \overline{V}, a, \eta, \rho \right\}$. Therefore, by calibrating these parameters values, we seek to obtain an evolution for the underlying asset that is consistent with the current prices of plain vanilla options.

In order to find the optimal parameter set we need to (i) define a measure to quantify the distance between model and market prices; and (ii) run an optimization scheme to determine the parameter values that minimize such distance. A simple and straightforward approach is to minimize the mean sum of squared differences

$$G(\Omega) = \sum_{i=1}^{N} \frac{1}{N} \left[ C_i^{\Omega}(K_i, T_i) - C_i^{Mkt}(K_i, T_i) \right]^2 \quad (4.1)$$

Where $C_i^{\Omega}(K_i, T_i)$ are the option values using the parameter set $\Omega$, and $C_i^{Mkt}(K_i, T_i)$ are the market observed option prices.

As shown in Bin (2007), the calibration process presents the problem that the objective function is not necessarily convex and may exhibit several local minima. This complicates the estimation of the optimal parameter set $\widehat{\Omega}$, since the solution attained by local optimization might be dependent on the initial guess $\Omega_0$. Therefore, a good initial guess might be critical and, even then, in some cases the convergence to the global optimum is not guaranteed.

The obvious solution is to employ global optimization. However, global optimizers generally lack the mathematical tractability of local ones, and also require substantially higher computational times. Since both methods have advantages and disadvantages, we will explore both approaches.

### 4.2 Local Optimization

When a function exhibits several minima, local optimizers face the problem that once a solution has been found, we cannot be sure whether such solution is the best available. In other words, we cannot distinguish if the solution is a local minimum or a global one, or consequently, if we have reached a local solution, there is no easy way to measure how far we are from the global one.

An alternative to tackle this problem is to define a criterion for acceptable solutions. If we select *a priori* which solutions can be deemed acceptable, we can at least ensure that any accepted solution will be consistent with our tolerance bounds. Conversely, if we found a non-acceptable solution, we can run the algorithm with a different starting point and keep searching for solutions that comply with our criteria.

In our tests, we will require that the difference between model and market prices falls on average within the observed bid-ask spreads. Therefore, we will consider the following set of acceptable solutions

$$\frac{1}{N}\sum_{i=1}^{N}\left|C_i^{\hat{\Omega}}(K_i,T_i)-C_i^{Mkt}(K_i,T_i)\right| \leq \frac{1}{2N}\sum_{i=1}^{N}\left[\text{bid}_i-\text{ask}_i\right] \quad (4.2)$$

where $C_i^{\hat{\Omega}}(K_i,T_i)$ are the model prices with the optimal parameter set $\hat{\Omega}$, $C_i^{Mkt}(K_i,T_i)$ are the mid-market option prices, and $\text{bid}_i / \text{ask}_i$ are the market observed bid and ask prices.

As a local optimizer we will use the Matlab *lsqnonlin* function (least-squares non-linear), which implements a trust-region reflective minimization algorithm[4]. In addition, we will also define lower and upper bounds for the optimal parameters. These thresholds are included in the calibration in order to avoid possible solutions that, while mathematically feasible, are not acceptable in an economic sense.

In particular, we will use the following bounds:

- **Long-term variance and initial variance:** Acceptable solutions for variance levels should take a possible value. However, given its mean-reversion, the volatility of most financial asset rarely reaches levels beyond 100%. Consequently, we will use bounds of 0 and 1 for both for $\bar{V}$ and $V_0$.
- **Correlation:** Statistical correlation takes values from -1 to 1. As previously mentioned, the correlation between volatility and stock prices tends to be negative. However, positive correlations might also be possible in particular cases. Therefore, the full range of acceptable solutions will be used in the calibration.
- **Volatility of variance:** Being a volatility, this parameter should exhibit positives values. However, the volatility of financial assets may change dramatically in short time periods (i.e. the volatility itself is very volatile). Consequently, high upper bounds are required for this parameter. In order to avoid potential restrictions, a broad set of solutions, from 0 to 5, will be used in the calibration.
- **Mean-reversion speed:** To ensure mean-reversion the parameter $a$ should take positive values (negative values will cause mean aversion). However, we have not found clear evidence regarding which upper value could be an appropriate bound. Consequently, instead of fixing an upper level, maximum values for $a$ will be dynamically set in the calibration as a by-product of the non-negativity constraint.
- **Non-negativity constraint:** In addition to the parameter bounds, another condition is required to ensure that the variance process in the Heston model does not reach zero or negative values. In this regard, Feller (1951) shows that a constraint $2a\bar{V}-\eta^2>0$ (generally known as the Feller condition) guarantees that the variance in a CIR process is always strictly positive[5].

---

[4] See Yuan (1999) for an overview on the use of trust-region algorithms for solving non-linear problems.

The option dataset that we use in the calibration are shown in Appendix D. Using the bounds described above, the implementation of the local calibration algorithm is shown in script 1 (*Heston_calibration_local.m*). In addition, function 5 (*costf.m*) provides the objective function required for script 1.

For dataset D1, the results obtained with local optimization are the following:

| $V_0$ | $\overline{V}$ | $\eta$ | $\rho$ | $a$ |
|-------|-------|-------|-------|-------|
| 0.0989 | 0.3407 | 0.7068 | -0.2949 | 0.7331 |

Using these results, the model predicted values and its comparison with the market prices are shown below:

| Option id. | Mid price | Model price | Difference(abs) | Within bid-ask? |
|---|---|---|---|---|
| 1 | 56.90 | 56.01 | 0.886 | YES |
| 2 | 36.30 | 35.57 | 0.728 | YES |
| 3 | 19.60 | 19.62 | 0.018 | YES |
| 4 | 9.45 | 9.26 | 0.185 | YES |
| 5 | 4.30 | 3.84 | 0.460 | NO |
| 6 | 63.20 | 63.26 | 0.059 | YES |
| 7 | 44.90 | 45.52 | 0.620 | NO |
| 8 | 30.55 | 31.07 | 0.519 | NO |
| 9 | 20.05 | 20.21 | 0.157 | YES |
| 10 | 12.50 | 12.69 | 0.188 | YES |
| 11 | 77.55 | 77.16 | 0.389 | YES |
| 12 | 61.45 | 61.87 | 0.420 | YES |
| 13 | 48.90 | 48.85 | 0.049 | YES |
| 14 | 38.45 | 38.10 | 0.349 | YES |
| 15 | 29.50 | 29.47 | 0.026 | YES |

As the table shows, the calibrated Heston model provides a good match for most traded options. 12 out of 15 options have a predicted value that falls within the observed bid-ask spread. In addition, when evaluated in terms of our acceptance criterion, the model's average distance from the mid-market price is 0.3369, which is lower than the average deviation in the bid-ask spreads (0.6933). The computational time required for the local calibration is 6.5 seconds.

However, the table also highlights a limitation of stochastic volatility models: these models may have problems to match the prices of out-of-the-money (OTM) options with short maturities (see, in particular, option n. 5[6]). More often than not, diffusion processes cannot generate the substantial underlying asset movements that are routinely implied by the prices of short-dated OTM options. Price jumps are generally perceived as one of the main drivers behind the high quotes for this type of options. Consequently, adding jumps to the underlying price process may be seen as a possible way forward which may improve the overall fit to market prices.

---

[5] This condition is particularly useful in certain Monte Carlo discretization schemes. In the calibration, the non-negativity constraint has been implemented by introducing an upper bound in the acceptable values of $2a\overline{V} - \eta^2$. Since $\overline{V}$ and $\eta^2$ have their own range of acceptable values, this condition implicitly restricts the acceptable values of $a$ to those that comply with the non-negativity constraint.

## 4.3 Global Optimization

The main advantage of global optimization is that it does not exhaust its search on the first minimum attained. Generally, global optimizers include stochastic movements in their search pattern, which make it possible to overcome local minimums and continue searching even if a potential solution has already been found.

However, the use of stochastic methods also entails certain drawbacks. The mathematical properties of these algorithms are less tractable than those of local (deterministic) ones. In addition, despite its name, their convergence to the global minimum is not guaranteed. In fact, since the exit sequence is determined stochastically, the algorithm might decide to terminate early and, in some cases, the solution attained might underperform a local search. All in all, even if global optimization is theoretically more powerful, when working with functions of unknown shape, it is not easy to establish *ex ante* which calibration method will perform better.

In order to test the results of global optimization we employ the Simulated Annealing framework (SA). This algorithm conducts a guided search, where new iterations are generated by taking into account the previous information but also introducing randomization. Initially, the algorithm starts with high tolerance for random shocks, and different regions are surveyed during the first phase. As a consequence, even if a minimum is found, the algorithm keeps searching for better solutions. As time evolves, the algorithm decreases its tolerance until it eventually settles in the best optimum attained.

In particular, we will use the Matlab function *asamin*, which was developed by Prof. Shinichi Sakata. This function implements an Adaptive Simulated Annealing (ASA), dynamically adjusting the tolerance for random shocks. The ASA framework has been shown by Goel and Stander (2009) to provide good results among a range of different global optimizers.

For comparability, we will use the same parameter bounds that we defined in section 4.2. The implementation of the *asamin* function is shown in script 2 (*Heston_calibration_global.m*), while the required cost function is implemented in function 6 (*costf_2.m*).

Running script 2, the optimal results obtained for dataset D1 are shown below:

| $V_0$ | $\bar{V}$ | $\eta$ | $\rho$ | $a$ |
|--------|--------|--------|--------|--------|
| 0.0983 | 0.2957 | 0.7544 | -0.2919 | 0.9626 |

---

[6] Individual contract details are included in Appendix D.

| Option id. | Mid price | Model price | Difference(abs) | Within bid-ask? |
|---|---|---|---|---|
| 1 | 56.90 | 56.05 | 0.853 | YES |
| 2 | 36.30 | 35.58 | 0.716 | YES |
| 3 | 19.60 | 19.59 | 0.008 | YES |
| 4 | 9.45 | 9.23 | 0.220 | YES |
| 5 | 4.30 | 3.83 | 0.470 | NO |
| 6 | 63.20 | 63.30 | 0.103 | YES |
| 7 | 44.90 | 45.55 | 0.647 | NO |
| 8 | 30.55 | 31.08 | 0.531 | NO |
| 9 | 20.05 | 20.21 | 0.165 | YES |
| 10 | 12.50 | 12.70 | 0.203 | YES |
| 11 | 77.55 | 77.13 | 0.416 | YES |
| 12 | 61.45 | 61.85 | 0.403 | YES |
| 13 | 48.90 | 48.85 | 0.055 | YES |
| 14 | 38.45 | 38.10 | 0.346 | YES |
| 15 | 29.50 | 29.48 | 0.017 | YES |

As can be seen, the optimal parameters values under ASA are slightly different to those of local calibration. However, there are not significant divergences in the overall results. Under global calibration 12 out of 15 model values are within the observed bid-ask spreads, and the average distance to the mid-market price is 0.3436. Therefore, the ASA solution is also acceptable according to our criterion and its quality is similar to the results obtained through Matlab's *lsqnonlin*. The main drawback of ASA is its substantially higher computational time (245.1 seconds in ASA vs 6.5 seconds in Matlab's *lsqnonlin*).

## 4.4 More Calibration Exercises

Based on dataset D1 both ASA and Matlab's *lsqnonlin* yield similar solutions. However, the complexity of multidimensional non-linear optimization makes it difficult to draw conclusions from a single comparison.

In order to obtain further evidence, we carried out two additional calibration exercises. First, we applied both methods to an option dataset which, *a priori*, should be easier to calibrate. In particular, all the options in dataset D2 have relatively broad bid-ask spreads and their implied volatilities are also relatively stable. Second, we also tested a potentially more challenging dataset (D3). In this case, the number of options was doubled and instruments with shorter maturities and divergent implied volatilities were included in the calibration.

The next table summarizes the calibration results for these datasets.

| Dataset | N. of options | Matlab's *lsqnonlin* | | | ASA (*asamin*) | | |
|---|---|---|---|---|---|---|---|
| | | Elapsed time | Within bid-ask | Average distance | Elapsed time | Within bid-ask | Average distance |
| D2 | 15 | 4.1 sec | 15 of 15 | 0.3903 | 258.0 sec | 15 of 15 | 0.4235 |
| D3 | 30 | 5.2 sec | 24 of 30 | 0.0197 | 562.4 sec | 24 of 30 | 0.0200 |

In dataset D2, both calibration methods produce good results. All the model predicted values are within the observed bid-ask spread. In terms of the distance from the mid-market prices, Matlab's *lsqnonlin* performs slightly better, with an average distance of 0.3903, against 0.4235 in ASA. In addition, as expected, the ASA algorithm takes substantially longer to reach the optimum.

Calibration gets more difficult in dataset D3. Although both methods provide acceptable solutions[7], the number of options within their observed bid-ask spread falls to 24 out of 30. However, even in these challenging conditions, the comparison between both methods exhibits a similar pattern, with Matlab's *lsqnonlin* reaching slightly better solution (average distance 0.0197 vs 0.0200) and ASA requiring significantly longer computing times.

Based on these exercises, we can conclude that Matlab's *lsqnonlin* provides better calibration results, and it also employs lower computational times. However, these results could be conditioned by an objective function that may not be complex enough to exploit the ASA strengths. In particular, since typically we do not know whether the objective function may exhibits several local minima, a conservative approach will be to run both calibration approaches. The drawback is, of course, that a global search might not necessarily improve the results provided by a local one. However, the advances in computing power and numerical methods keep reducing the time required for global calibration. In our exercises, the running time of ASA was lower than 10 minutes, which for many practical applications makes it worth testing for potentially better solutions.

**Script 1: Heston local calibration using Matlab's lsqnonlin (Heston_calibration_local.m)**

```
% Heston calibration, local optimization (Matlab's lsqnonlin)

% Input on data.txt
% Data = [So, t, k, r, mid price, bid, ask]
clear all
global data; global cost; global finalcost;
load data.txt

% Initial parameters and parameter bounds
% Bounds [v0, Vbar,  vvol, rho, 2*a*vbar - vvol^2]
% Last bound include non-negativity constraint and bounds for mean-reversion
x0 = [.5,.5,1,-0.5,1];
lb = [0, 0, 0, -1, 0];
ub = [1, 1, 5, 1, 20];

% Optimization: calls function costf.m:
tic;
x = lsqnonlin(@costf,x0,lb,ub);
toc;

% Solution:
```

---

[7] The average observed deviation in the market bid-ask spreads is 0.0559.

```
Heston_sol = [x(1), x(2), x(3), x(4), (x(5)+x(3)^2)/(2*x(2))]
x
min = finalcost
```

**Matlab Function 5: Cost function for local calibration (costf.m)**

```
function [cost] = costf(x)
global data; global finalcost;

% Compute individual differences
% Sum of squares performed by Matlab's lsqnonlin
for i=1:length(data)
cost(i)= data(i,5) - call_heston_cf(data(i,1),x(1), x(2), (x(5)+x(3)^2)/(2*x(2)), x(3), data(i,4), x(4), data(i, 2),
data(i,3));
end

% Show final cost
finalcost =sum(cost)^2
end
```

**Script 2: Heston global calibration using ASA (Heston_calibration_global.m)**

```
% Heston calibration, global optimization (asamin)

% Input on data.txt
% Data = [So, t, k, r, mid price, bid, ask]
clear all
global data; global cost; global finalcost;
load data.txt

% Initial parameters and parameter bounds
% Bounds [v0, Vbar,  vvol, rho, 2*a*vbar - vvol^2]
% Last bound include non-negativity constraint and bounds for mean-reversion
x0 = [.5,.5,1,-0.5,5];
lb = [0, 0, 0, -1, 0];
ub = [1, 1, 6, 1, 20];

% Optimization: calls function costf_2.m:
asamin('set', 'test_in_cost_func', 0);
xtype = [-1;-1;-1;-1;-1];
tic;
[f, x_opt, grad, hessian, state] = asamin ('minimize','costf_2' ,x0',lb',ub', xtype)
toc;

% Solution:
Heston_sol = [x(1), x(2), x(3), x(4), (x(5)+x(3)^2)/(2*x(2))]
x
min = finalcost
```

**Matlab Function 6: Cost function for global calibration (costf_2.m)**

```
function [cost flag] = costf_2(x)
global data; global finalcost; global cost; global cost_i;

% Compute individual differences
for i=1:length(data)
cost_i(i)= data(i,5) - call_heston_cf(data(i,1),x(1), x(2), (x(5)+x(3)^2)/(2*x(2)), x(3), data(i,4), x(4), data(i,
2), data(i,3));
end

% Compute sum of squared differences
cost = sum(cost_i.^2);

% Show final cost and current solution
finalcost =sum(cost)
flag = 1;
Heston_sol = [x(1), x(2), x(3), x(4), (x(5)+x(3)^2)/(2*x(2))]
end
```

## 5. Conclusion

Stochastic volatility models tackle one of the most restrictive hypotheses of the BSM framework, which assumes that volatility remains constant during the option´s life. However, by observing financial markets it becomes apparent that volatility may change dramatically in short-time periods and its behavior is clearly not deterministic.

Among stochastic volatility models, the Heston model presents two main advantages. First, it models an evolution of the underlying asset which can take into account the asymmetry and excess kurtosis that are typically observed (and expected) in financial asset returns. Second, it provides closed-form solutions for the pricing of European options.

Availability of closed-form valuation formulas is particularly important for the calibration process. In our tests, although the objective function is not necessarily convex, both local and global optimization methods provide reasonable results within a relatively short timeframe. However, in cases where the objective function may exhibit several local minima, local optimization may underperform a global search. Once the model parameters have been calibrated to fit market prices, the Heston dynamics can be used to price other products that are not actively traded in the market.

Following these results there are also two possible areas of further work. First, before using the calibrated model to price exotic products, a discretization scheme will be typically required in order to obtain more granular information regarding the underlying asset dynamics during the product´s life. This can be achieved, in most practical cases, by implementing a Monte Carlo simulation scheme.

Second, a step further will be to include discontinuous jumps in the underlying asset evolution. Adding jumps to stochastic volatility entails higher complexity, but also provides a potentially more realistic framework. Most jump models follow a characteristic function approach whose implementation is similar to the one described here. Therefore, for interested readers, we hope that the explanations provided in this paper may help them to connect the dots in their next mathematical journey.

## Appendix A:  Derivation of $\Pi_1$ and $\Pi_2$

The proof is divided in two parts. In the first one we derive $\Pi_1$ and $\Pi_2$ based on the relationship between the cumulative density function (CDF) of a random variable $X$ and its characteristic function

$$F(x) = \frac{1}{2} - \frac{1}{\pi} \int_0^\infty \text{Re}\left[ \frac{e^{-iwx}\psi_X(w)}{iw} \right] dw \quad \text{(A.1)}$$

The second part is devoted to prove (A.1).

***

For the first part we follow the reasoning in Chourdakis (2008). We start with the value of European call with maturity date T and strike K. In a risk-neutral context, the call value at t = 0 is given by

$$C_0 = e^{-rT} E_Q\left[ \max(S_T - K, 0) \right] \qquad \text{(A.2)}$$

Using x = $\ln(S_T)$ and expanding (A.2) we get an expression for the European call value that is similar to the definition in terms of $\Pi_1$ and $\Pi_2$ that we used in (2.5)

$$\begin{aligned}
C_0 &= e^{-rT} \int_{\log K}^\infty (e^x - K) f(x) dx \\
&= e^{-rT} \left( \int_{\log K}^\infty e^x f(x) dx - K \int_{\log K}^\infty f(x) dx \right) \quad \text{(A.3)} \\
&= e^{-rT} I_1 - e^{-rT} K I_2
\end{aligned}$$

For a given call option, by comparing equations (2.5) and (A.3) it can be seen that $\Pi_2$ should be equal to $I_2$, while $\Pi_1$ should be equal to $I_1 e^{-rT} / S_0$. The second integral $I_2$ is simply the probability of the log-stock price finishing above the log-strike. Therefore, by applying the relationship in (A.1), this probability can be obtained in terms of the characteristic function of $\ln(S_T)$ as follows

$$\begin{aligned}
\Pi_2 = I_2 &= P(\ln S_T > \ln K) \\
&= 1 - P(\ln S_T \le \ln K) \\
&= \frac{1}{2} + \frac{1}{\pi} \int_0^\infty \text{Re}\left[ \frac{e^{-iw\ln(K)}\psi_{\ln S_T}(w)}{iw} \right] dw
\end{aligned}$$

which is the definition of $\Pi_2$ that we presented in (2.7).

To derive $\Pi_1$, we multiply and divide the first integral $I_1$ by the term $\int_{-\infty}^\infty e^x f(x) dx$, which is also equal, in a risk-neutral context, to the capitalized spot price (i.e. $e^{rT} S_0$)

$$I_1 = \int_{\log K}^{\infty} e^x f(x) dx = \frac{\int_{\log K}^{\infty} e^x f(x) dx}{\int_{-\infty}^{\infty} e^x f(x) dx} \int_{-\infty}^{\infty} e^x f(x) dx$$

$$= g(x) e^{rT} S_0$$

Working on the fraction above, we obtain an alternative integral expression for $g(x)$ as follows

$$g(x) = \frac{\int_{\log K}^{\infty} e^x f(x) dx}{\int_{-\infty}^{\infty} e^x f(x) dx} = \int_{\log K}^{\infty} \left( \frac{e^x f(x)}{\int_{-\infty}^{\infty} e^x f(x) dx} \right) dx = \int_{\log K}^{\infty} \left( f^*(x) \right) dx$$

Therefore, the first integral $I_1$ can be also expressed as

$$I_1 = e^{rT} S_0 \int_{\log K}^{\infty} f^*(x) dx$$

Since $f^*(x)$ is, by construction, between 0 and 1, its Fourier transforms is given by

$$\psi^*(w) = \int_{-\infty}^{\infty} e^{iwx} f^*(x) dx = \frac{\psi(w-i)}{\psi(-i)}$$

Consequently, using again the relationship in (A.1)

$$I_1 = e^{rT} S_0 \left( \frac{1}{2} + \frac{1}{\pi} \int_0^{\infty} \text{Re} \left[ \frac{e^{-iw\ln(K)} \psi_{\ln S_T}(w-i)}{iw \psi_{\ln S_T}(-i)} \right] dw \right)$$

Finally, since $\Pi_1 = I_1 e^{-rT} / S_0$, the expression for $\Pi_1$ simplifies to:

$$\Pi_1 = \frac{1}{2} + \frac{1}{\pi} \int_0^{\infty} \text{Re} \left[ \frac{e^{-iw\ln(K)} \psi_{\ln S_T}(w-i)}{iw \psi_{\ln S_T}(-i)} \right] dw$$

which is the definition of $\Pi_1$ that we used in (2.6).

$$***$$

The second part follows the reasoning in Kendall, Stuart and Ord (1994) and Wu (2007).

First we start with the integral

$$I = \int_0^{\infty} \frac{e^{iwx} \psi_X(-w) - e^{-iwx} \psi_X(w)}{iw} dw$$

Replacing each characteristic function by its integral form, the expression above becomes

$$I = \int_0^\infty \frac{e^{iwx}\int_{-\infty}^\infty e^{-iwz}dF(z) - e^{-iwx}\int_{-\infty}^\infty e^{iwz}dF(z)}{iw}dw$$

$$= \int_0^\infty \int_{-\infty}^\infty \frac{e^{iwx}e^{-iwz} - e^{-iwx}e^{iwz}}{iw}dF(z)dw$$

$$= \int_0^\infty \int_{-\infty}^\infty \frac{e^{iw(x-z)} - e^{-iw(x-z)}}{iw}dF(z)dw$$

Next, considering Euler's equality $\sin(\theta) = (e^{i\theta} - e^{-i\theta})/2i$, and using $\theta = w(x-z)$, it can be seen that $2\sin w(x-z) = (e^{iw(x-z)} - e^{-iw(x-z)})/i$.

Therefore, applying Fubini's theorem and the fact that $\lim_{n\to\infty}\int_0^n \sin(\delta t)/t\,dt = \pi/2\,\mathrm{sgn}(\delta)$ the integral simplifies to

$$I = \int_0^\infty \int_{-\infty}^\infty \frac{2\sin w(x-z)}{w}dF(z)dw$$

$$= \int_{-\infty}^\infty \int_0^\infty \frac{2\sin w(x-z)}{w}dw\,dF(z)$$

$$= \int_{-\infty}^\infty \pi\,\mathrm{sgn}(x-z)dF(z) =$$

$$= \pi\big[(F(x) + 0 - (1 - F(x)))\big]$$

$$= \pi\big[2F(x) - 1\big]$$

Consequently, solving for $F(x)$ and then substituting $I$ by its original definition yields

$$F(x) = \frac{1}{2} + \frac{1}{2\pi}I$$

$$= \frac{1}{2} + \frac{1}{2\pi}\int_0^\infty \frac{e^{iwx}\psi_X(-w) - e^{-iwx}\psi_X(w)}{iw}dw$$

Finally, since the density of $X$ is a real-valued function, using the properties of Fourier transforms, $\psi_X(w)$ has conjugate symmetry and $[\psi_X(w) + \psi_X(-w)]/2 = \mathrm{Re}[\psi_X(w)]$. Therefore, the CDF of $X$ can also be expressed as

$$F(x) = \frac{1}{2} + \frac{1}{\pi}\int_0^\infty \frac{e^{iwx}\psi_X(-w) - e^{-iwx}\psi_X(w)}{2iw}dw$$

$$= \frac{1}{2} - \frac{1}{\pi}\int_0^\infty \mathrm{Re}\left[\frac{e^{-iwx}\psi_X(w)}{iw}\right]dw$$

which is the definition of $F(x)$ that we used in (A.1).

## Appendix B: Equivalence of our approach to Gatheral (2006)

The analysis of the Heston call value in Gatheral (2006) is based on the process $x_T = \ln(F_T / K)$, where $F_T$ is the forward price of the underlying asset at the maturity date T. Consequently, its derivation focus on the future value of the European call at time $t = T$

$$C_T^{Ga} = K(e^{x_T}\Pi_1 - \Pi_2) \quad \text{(B.1)}$$

rather than its value today. However, taking into account that the forward price is $F_T = S_0 e^{rT}$, equation (B.1) becomes

$$C_T^{Ga} = K(e^{\ln(S_0 e^{rT} / K)}\Pi_1 - \Pi_2)$$
$$= S_0 e^{rT}\Pi_1 - K\Pi_2$$

and calculating the present value of the expression above (i.e. multiplying by $e^{-rT}$ in a risk neutral context) yields the probabilistic definition of the European call value that we used through the paper

$$C_0 = S_0\Pi_1 - e^{-rT}K\Pi_2$$

Next, we need to show that the definitions that we used for $\Pi_1$ and $\Pi_2$ are equivalents to those provided by Gatheral (2006). Regarding $\Pi_2$ (i.e. probability of the final log-stock price being greater than the log-strike), the result provided in Gatheral's is given by

$$\Pi_2^{Ga} = \frac{1}{2} + \frac{1}{\pi}\int_0^\infty \text{Re}\left[\frac{e^{C(T,w)\bar{v} + D(T,w)v_0 + iwx}}{iw}\right]dw \quad \text{(B.2)}$$

where $x = \ln(F_T / K)$, and $C(T,w)$, $D(T,w)$ are defined in the same terms that we used in (3.3). Expanding Gatheral's result we obtain

$$\Pi_2^{Ga} = \frac{1}{2} + \frac{1}{\pi}\int_0^\infty \text{Re}\left[\frac{e^{iw\ln(F_T / K)}e^{C(T,w)\bar{v} + D(T,w)v_0}}{iw}\right]dw$$
$$= \frac{1}{2} + \frac{1}{\pi}\int_0^\infty \text{Re}\left[\frac{e^{iw\ln(F_T)}e^{-iw\ln(K)}e^{C(T,w)\bar{v} + D(T,w)v_0}}{iw}\right]dw$$
$$= \frac{1}{2} + \frac{1}{\pi}\int_0^\infty \text{Re}\left[\frac{e^{-iw\ln(K)}e^{C(T,w)\bar{v} + D(T,w)v_0 + iw\ln(F_T)}}{iw}\right]dw$$

And recalling that, at time $t = T$, the characteristic function of the Heston model that we used in (3.3) is precisely

$$\psi_{\ln(S_T)}^{Heston}(w) = e^{[C(T,w)\bar{v} + D(T,w)v_0 + iw\ln(S_0 e^{rT})]}$$

the expression for $\Pi_2^{Ga}$ becomes

$$\Pi_2^{Ga} = \frac{1}{2} + \frac{1}{\pi} \int_0^\infty \mathrm{Re}\left[\frac{e^{-iw\ln(K)}\psi_{\ln S_T}(w)}{iw}\right] dw$$

which is the definition of $\Pi_2$ that we have used through the paper.

A similar approach can be used to show the equivalence of (2.6) to the expression for $\Pi_1^{Ga}$ provided in Gatheral (2006).

## Appendix C: Risk Neutrality in the Heston model

In order to understand the use of risk neutrality we first state the main result and then we prove it.

**Main result**

We start with the Heston dynamics under the physical measure P

$$dS_t = \mu S_t dt + \sqrt{V_t} S_t dW_t^{P,1}$$
$$dV_t = a^P(\bar{V}^P - V_t)dt + \eta\sqrt{V_t}dW_t^{P,2}$$
$$dW_t^{P,1}dW_t^{P,2} = \rho dt$$

and we seek to obtain a risk-neutral evolution where $E_t^Q(dS_t / S_t) = rdt$. As we show below, using the multidimensional Girsanov's theorem and making appropriate choices, the Heston dynamics under the risk-neutral measure Q can be expressed as

$$dS_t = rS_t dt + \sqrt{V_t} S_t dW_t^{Q,1}$$
$$dV_t = a^Q(\bar{V}^Q - V_t)dt + \eta\sqrt{V_t}dW_t^{Q,2}$$
$$dW_t^{Q,1}dW_t^{Q,2} = \rho dt$$

where $a^Q = a^P + \gamma$, $\bar{V}^Q = \frac{a^P\bar{V}}{a^P + \gamma}$ and $\gamma$ is a parameter linked to the price of volatility risk.

Therefore, the Heston dynamics under the risk-neutral measure exhibit a similar pattern to that of the physical measure, but with a variance process that is defined by the parameters $a^Q$ and $\bar{V}^Q$ instead of $a^P$ and $\bar{V}^P$. A remarkable feature is that $a^Q$ and $\bar{V}^Q$ already incorporate the impact of the volatility risk premium $\gamma$. Consequently, when calibrating the risk-neutral model to market prices, we can directly solve for $a^Q$ and $\bar{V}^Q$, and we will not need to estimate $\gamma$ explicitly.

In section 3, for simplicity, we omitted the Q superscripts. However, it should be noted that the values for $a$ and $\bar{V}$ that we used through the paper are the risk-neutral ones (i.e. $a^Q$ and $\bar{V}^Q$), and not those under the physical measure. The use of risk-neutral dynamics is justified when all the risks related to holding options can be hedged away. Within the Heston model, there are two sources of uncertainty: the underlying asset movements and the volatility movements. The first risk source can be hedged away implementing a delta-hedging strategy in similar terms to those of the BSM framework.

However, in order to hedge the volatility risk, a liquid market for volatility related contracts is needed. Consequently, the use of risk-neutral pricing is conditioned by the assumption of perfect hedging. If hedging is not possible, we might need to go back to the dynamics under the physical measure, which requires different models and hypothesis in order to estimate the appropriate risk premiums and the corresponding real-world distribution.

**Proof**

We start again with the Heston dynamics under the physical measure

$$dS_t = \mu S_t dt + \sqrt{V_t} S_t dW_t^{P,1}$$
$$dV_t = a^P(\bar{V}^P - V_t)dt + \eta\sqrt{V_t} dW_t^{P,2} \quad \text{(C.1)}$$
$$dW_t^{P,1} dW_t^{P,2} = \rho dt$$

where the discounted underlying price is a martingale under P.

To obtain the risk-neutral dynamics we should find an equivalent martingale measure (EMM) where the process $dS_t/S_t$ has a drift of $rdt$. To achieve this we perform a change of probability measure using Girsanov's theorem. In particular, we define a new EMM through the Radon-Nikodym derivative:

$$\left.\frac{dQ}{dP}\right|_t = M_t$$

where $M_t$ is an exponential martingale of the form

$$M_t = \exp\left\{\int_t^T C_s dW_s^{P,1} - \frac{1}{2}\int_t^T C_s^2 ds \ + \ \int_t^T D_s dW_s^{P,2} - \frac{1}{2}\int_t^T D_s^2 ds\right\}$$

and it is the solution of the SDE

$$\frac{dM_t}{M_t} = C_t dW^{P,1} + D_t dW^{P,2}$$

with initial value $M_0 = 1$.

Since we are working with EMMs, the expectation of a given stochastic process Z under the new measure Q can be computed as

$$E_t^Q(Z) = E_t^P(M_t Z)$$

Therefore, if we consider the expectation of infinitesimal increments

$$E_t^Q(dZ) = E_t^P\left(\frac{M_t + dMt}{Mt}dZ\right) = E_t^P\left[\left(1 + \frac{dMt}{Mt}\right)dZ\right] = E_t^P dZ + E_t^P(C_t dW^{P,1} + D_t dW^{P,2})dZ$$

Using the equation above, we can compute the drift and volatility for the process $dS_t/S_t$ under Q

$$E_t^Q\left(\frac{dS_t}{S_t}\right) = E_t^P\left(\frac{dS_t}{S_t}\right) + E_t^P\left((C_t dW^{P,1} + D_t dW^{P,2})\frac{dS_t}{S_t}\right)$$

$$= E_t^P(\mu dt + \sqrt{V_t} dW_t^{P,1}) + E_t^P\left((C_t dW^{P,1} + D_t dW^{P,2})(\mu dt + \sqrt{V_t} dW_t^{P,1})\right)$$

$$= E_t^P(\mu dt) + E_t^P(\sqrt{V_t} dW_t^{P,1}) + E_t^P(C_t \mu dW^{P,1} dt) + E_t^P(C_t \sqrt{V_t}(dW_t^{P,1})^2) +$$

$$+ E_t^P(D_t \mu dW^{P,2} dt) + E_t^P(D_t \sqrt{V_t} dW^{P,2} dW_t^{P,1})$$

$$= \mu dt + 0 + 0 + C_t \sqrt{V_t} dt + 0 + \rho D_t \sqrt{V_t} dt$$

$$= (\mu + C_t \sqrt{V_t} + D_t \sqrt{V_t} \rho) dt$$

$$E_t^Q\left(\frac{dS_t}{S_t}\right)^2 = E_t^P\left[\left(1 + C_t dW^{P,1} + D_t dW^{P,2}\right)\left(\mu dt + \sqrt{V_t} dW_t^{P,1}\right)\right]^2$$

$$= E_t^P\left[V_t(dW_t^{P,1})^2\right]$$

$$= V_t dt$$

where we expanded the initial expressions and we used the fact that Weiner processes are distributed as $N(0, \sqrt{t})$ and, consequently, $E(dW_t^{P,1} dW_t^{P,2}) = \rho dt$. We also used the basic rules of stochastic calculus $E(dW_t) = 0$; $E(dW_t dt) = 0$; $E(dt^2) = 0$ and $E\left[(dW_t)^2\right] = dt$.

Similarly, the drift and volatility for $dV_t$ can be computed as

$$E_t^Q\left(dV_t\right) = E_t^P\left(dV_t\right) + E_t^P\left((C_t dW^{P,1} + D_t dW^{P,2}) dV_t\right)$$

$$= E_t^P(a^P(\bar{V}^P - V_t) dt) + E_t^P\left((C_t dW^{P,1} + D_t dW^{P,2})\left[a^P(\bar{V}^P - V_t) dt + \eta \sqrt{V_t} dW_t^{P,2}\right]\right)$$

$$= a^P(\bar{V}^P - V_t) dt + E_t^P\left[C_t \eta \sqrt{V_t} dW^{P,1} dW^{P,2}\right] + E_t^P\left[D_t \eta \sqrt{V_t}(dW^{P,2})^2\right]$$

$$= \left[a^P(\bar{V}^P - V_t) + \rho C_t \eta \sqrt{V_t} + D_t \eta \sqrt{V_t}\right] dt$$

$$E_t^Q\left(dV_t\right)^2 = E_t^P\left[\left(1 + C_t dW^{P,1} + D_t dW^{P,2}\right)\left(a^P(\bar{V}^P - V_t) dt + \eta \sqrt{V_t} dW_t^{P,2}\right)\right]^2$$

$$= E_t^P\left[\eta^2 V_t(dW_t^{P,2})^2\right]$$

$$= \eta^2 V_t dt$$

Now, in order select the desired EMM, we impose the restriction

$$E_t^Q\left(\frac{dS_t}{S_t}\right) = r dt$$

which gives us the equation $(\mu + C_t \sqrt{V_t} + D_t \sqrt{V_t} \rho) dt = r dt$. Rearranging terms we obtain the following relationship, which defines the market price of risk

$$C_t + \rho D_t = -\frac{\mu - r}{\sqrt{V_t}}$$

Additionally, we need to set the drift for the volatility process. In this case, an appropriate choice is

$$E_t^Q \left( dV_t \right) = \left[ a^P (\bar{V}^P - V_t) - \gamma V_t \right] dt$$

where $\gamma$ is a parameter related to the price of volatility risk. This constraint gives us the equation $[a^P (\bar{V}^P - V_t) + \rho C_t \eta \sqrt{V_t} + D_t \eta \sqrt{V_t}] dt = [a^P (\bar{V}^P - V_t) - \gamma V_t] dt$, which defines the price of volatility risk

$$\rho C_t + D_t = -\frac{\gamma \sqrt{V_t}}{\eta}$$

Considering the properties of EMMs, the multidimensional Girsanov´s theorem tells us that the Weiner processes under the new measure Q are

$$W_t^{Q,1} = W_t^{P,1} + \frac{\mu - r}{\sqrt{V_t}} t$$

$$W_t^{Q,2} = W_t^{P,2} + \frac{\gamma \sqrt{V_t}}{\eta} t$$

Therefore, rearranging terms and substituting on the initial dynamics we get

$$dS_t = \mu S_t dt + \sqrt{V_t} S_t dW_t^{P,1}$$

$$= \mu S_t dt + \sqrt{V_t} S_t d\left( W_t^{Q,1} - \frac{\mu - r}{\sqrt{V_t}} t \right)$$

$$= \mu S_t dt + \sqrt{V_t} S_t dW_t^{Q,1} - \sqrt{V_t} S_t \frac{\mu - r}{\sqrt{V_t}} dt$$

$$= r S_t dt + \sqrt{V_t} S_t dW_t^{Q,1}$$

$$dV_t = a^P (\bar{V}^P - V_t) dt + \eta \sqrt{V_t} dW_t^{P,2}$$

$$= a^P (\bar{V}^P - V_t) dt + \eta \sqrt{V_t} d\left( W_t^{Q,2} - \frac{\gamma \sqrt{V_t}}{\eta} t \right)$$

$$= a^P (\bar{V}^P - V_t) dt + \eta \sqrt{V_t} dW_t^{Q,2} - \eta \sqrt{V_t} \frac{\gamma \sqrt{V_t}}{\eta} dt$$

$$= \left( a^P \bar{V}^P - a^P V_t - \gamma V_t \right) dt + \eta \sqrt{V_t} dW_t^{Q,2}$$

and if we introduce the notation $a^Q = a^P + \gamma$ and $\bar{V}^Q = \frac{a^P \bar{V}}{a^P + \gamma}$, the process $dV_t$ becomes

$$dV_t = a^Q (\bar{V}^Q - V_t) dt + \eta \sqrt{V_t} dW_t^{Q,2}$$

Finally, the correlation condition $dW_t^{P,1}dW_t^{P,2} = \rho dt$ is equivalent to require $E(dW_t^{P,1}dW_t^{P,2}) = \rho dt$. And considering the relationship between the Weiner processes under the physical and the risk-neutral measure we get

$$\begin{aligned}
\rho dt &= E(dW_t^{P,1}dW_t^{P,2}) \\
&= E\left[ d\left( W_t^{Q,1} + \frac{\mu - r}{\sqrt{V_t}}t \right) d\left( W_t^{Q,2} + \frac{\gamma\sqrt{V_t}}{\eta}t \right) \right] \\
&= E\left( dW_t^{Q,1}dW_t^{Q,2} \right)
\end{aligned}$$

where we have used again the stochastic calculus rules $E(dW_t) = 0$; $E(dW_t dt) = 0$ and $E(dt^2) = 0$.

## Appendix D: Datasets used for Calibration

**Dataset D1:** 15 options (3 maturities, 5 strikes).

| Spot | Maturity | Strike | Interest rate | Mid | Bid | Ask |
|------|----------|--------|---------------|------|------|------|
| 328.29 | 0.1753424 | 275 | 0.000553778 | 56.9 | 55.5 | 58.3 |
| 328.29 | 0.1753424 | 300 | 0.000553778 | 36.3 | 35.0 | 37.6 |
| 328.29 | 0.1753424 | 325 | 0.000553778 | 19.6 | 19.3 | 19.9 |
| 328.29 | 0.1753424 | 350 | 0.000553778 | 9.45 | 9.2 | 9.7 |
| 328.29 | 0.1753424 | 375 | 0.000553778 | 4.3 | 4.1 | 4.5 |
| 328.29 | 0.4246575 | 275 | 0.000659467 | 63.2 | 61.7 | 64.7 |
| 328.29 | 0.4246575 | 300 | 0.000659467 | 44.9 | 44.4 | 45.4 |
| 328.29 | 0.4246575 | 325 | 0.000659467 | 30.55 | 30.2 | 30.9 |
| 328.29 | 0.4246575 | 350 | 0.000659467 | 20.05 | 19.7 | 20.4 |
| 328.29 | 0.4246575 | 375 | 0.000659467 | 12.5 | 12.2 | 12.8 |
| 328.29 | 0.9232876 | 275 | 0.000850338 | 77.55 | 76.1 | 79.0 |
| 328.29 | 0.9232876 | 300 | 0.000850338 | 61.45 | 60.8 | 62.1 |
| 328.29 | 0.9232876 | 325 | 0.000850338 | 48.9 | 48.1 | 49.7 |
| 328.29 | 0.9232876 | 350 | 0.000850338 | 38.45 | 37.9 | 39.0 |
| 328.29 | 0.9232876 | 375 | 0.000850338 | 29.5 | 29.0 | 30.0 |

Call options written on Biogen Idec (Nasdaq: BIIB). Market data observed on February 14, 2014

**Dataset D2:** 15 options (3 maturities, 5 strikes)

| Spot | Maturity | Strike | Interest rate | Mid | Bid | Ask |
|------|----------|--------|---------------|------|------|------|
| 1313.67 | 0.3972602 | 1200 | 0.000697973 | 160.15 | 158.6 | 161.7 |
| 1313.67 | 0.3972602 | 1250 | 0.000697973 | 127.25 | 125.6 | 128.9 |
| 1313.67 | 0.3972602 | 1300 | 0.000697973 | 99.15 | 98.0 | 100.3 |
| 1313.67 | 0.3972602 | 1350 | 0.000697973 | 75.25 | 73.8 | 76.7 |
| 1313.67 | 0.3972602 | 1400 | 0.000697973 | 55.6 | 54.4 | 56.8 |
| 1313.67 | 0.8958904 | 1200 | 0.000853821 | 211.1 | 209.4 | 212.8 |
| 1313.67 | 0.8958904 | 1250 | 0.000853821 | 182.25 | 180.6 | 183.9 |
| 1313.67 | 0.8958904 | 1300 | 0.000853821 | 156.35 | 155.0 | 157.7 |
| 1313.67 | 0.8958904 | 1350 | 0.000853821 | 132.2 | 130.3 | 134.1 |
| 1313.67 | 0.8958904 | 1400 | 0.000853821 | 111.55 | 110.2 | 112.9 |
| 1313.67 | 1.8904109 | 1200 | 0.002228013 | 286 | 284.2 | 287.8 |
| 1313.67 | 1.8904109 | 1250 | 0.002228013 | 259.75 | 257.8 | 261.7 |
| 1313.67 | 1.8904109 | 1300 | 0.002228013 | 235.3 | 233.2 | 237.4 |
| 1313.67 | 1.8904109 | 1350 | 0.002228013 | 213.05 | 211.2 | 214.9 |
| 1313.67 | 1.8904109 | 1400 | 0.002228013 | 192.2 | 190.4 | 194.0 |

Call options written on The Priceline Group (Nasdaq: PCLN). Market data observed on February 24, 2014

**Dataset D3:** 30 options (6 maturities, 5 strikes)

| Spot | Maturity | Strike | Interest rate | Mid | Bid | Ask |
|------|----------|--------|---------------|-----|-----|-----|
| 39.63 | 0.0493150 | 36 | 0.000631752 | 3.75 | 3.7 | 3.8 |
| 39.63 | 0.0493150 | 38 | 0.000631752 | 2.145 | 2.13 | 2.16 |
| 39.63 | 0.0493150 | 40 | 0.000631752 | 1.035 | 1.02 | 1.05 |
| 39.63 | 0.0493150 | 42 | 0.000631752 | 0.435 | 0.42 | 0.45 |
| 39.63 | 0.0493150 | 44 | 0.000631752 | 0.17 | 0.16 | 0.18 |
| 39.63 | 0.1260273 | 36 | 0.000707312 | 4.3 | 4.25 | 4.35 |
| 39.63 | 0.1260273 | 38 | 0.000707312 | 2.91 | 2.89 | 2.93 |
| 39.63 | 0.1260273 | 40 | 0.000707312 | 1.85 | 1.84 | 1.86 |
| 39.63 | 0.1260273 | 42 | 0.000707312 | 1.095 | 1.08 | 1.11 |
| 39.63 | 0.1260273 | 44 | 0.000707312 | 0.615 | 0.61 | 0.62 |
| 39.63 | 0.3753424 | 36 | 0.000734416 | 5.55 | 5.5 | 5.6 |
| 39.63 | 0.3753424 | 38 | 0.000734416 | 4.35 | 4.3 | 4.4 |
| 39.63 | 0.3753424 | 40 | 0.000734416 | 3.35 | 3.3 | 3.4 |
| 39.63 | 0.3753424 | 42 | 0.000734416 | 2.55 | 2.53 | 2.57 |
| 39.63 | 0.3753424 | 44 | 0.000734416 | 1.92 | 1.9 | 1.94 |
| 39.63 | 0.6246575 | 36 | 0.000796417 | 6.475 | 6.4 | 6.55 |
| 39.63 | 0.6246575 | 38 | 0.000796417 | 5.35 | 5.3 | 5.4 |
| 39.63 | 0.6246575 | 40 | 0.000796417 | 4.4 | 4.35 | 4.45 |
| 39.63 | 0.6246575 | 42 | 0.000796417 | 3.6 | 3.55 | 3.65 |
| 39.63 | 0.6246575 | 44 | 0.000796417 | 2.92 | 2.89 | 2.95 |
| 39.63 | 0.8739726 | 35 | 0.000882340 | 7.775 | 7.7 | 7.85 |
| 39.63 | 0.8739726 | 37 | 0.000882340 | 6.675 | 6.6 | 6.75 |
| 39.63 | 0.8739726 | 40 | 0.000882340 | 5.25 | 5.2 | 5.3 |
| 39.63 | 0.8739726 | 42 | 0.000882340 | 4.425 | 4.35 | 4.5 |
| 39.63 | 0.8739726 | 45 | 0.000882340 | 3.425 | 3.35 | 3.5 |
| 39.63 | 1.8684931 | 35 | 0.002280481 | 10.125 | 9.95 | 10.3 |
| 39.63 | 1.8684931 | 37 | 0.002280481 | 9.2 | 9.05 | 9.35 |
| 39.63 | 1.8684931 | 40 | 0.002280481 | 7.85 | 7.75 | 7.95 |
| 39.63 | 1.8684931 | 42 | 0.002280481 | 7.1 | 7.0 | 7.2 |
| 39.63 | 1.8684931 | 45 | 0.002280481 | 6.1 | 5.95 | 6.25 |

Call options written on Yahoo (Nasdaq: YHOO). Market data observed on March 4, 2014.